\documentstyle[mprocl]{article}

\input epsf

\input{psfig}

\def \fun#1#2{\lower3.6pt\vbox{\baselineskip0pt\lineskip.9pt
\ialign{$\mathsurround=0pt#1\hfil##\hfil$\crcr#2\crcr\sim\crcr}}}
\begin{document}

\title{CAN WE APPLY STATISTICAL LAWS TO SMALL SYSTEMS?\\
      THE CERIUM ATOM.} 

\author{I. V. PONOMAREV$^{a}$, V. V. FLAMBAUM,
  A. A. GRIBAKINA and G. F. GRIBAKIN}

\address{School of Physics, The University of New South Wales,
Sydney 2052, Australia}

\maketitle\abstracts{
It is shown that statistical mechanics is applicable to quantum
systems with finite numbers of particles, such as complex atoms, atomic
clusters, etc., where the residual two-body interaction is sufficiently
strong. This interaction mixes the unperturbed shell-model
basis states and produces ``chaotic'' many-body eigenstates. As a result,
an interaction-induced equilibrium emerges in the system, and temperature
can be introduced. However, the interaction between the particles and
their finite number can lead to prominent deviations of the equilibrium
occupation numbers distribution from the Fermi-Dirac shape. For
example, this takes place in the cerium atom with four valence electrons,
which was used to compare the theory with realistic numerical
calculations.}

\vspace*{-9pt}
%*********************************************************************
{\bf 1. Introduction.}
Statistical behaviour is usually established in the limit of
a large number of particles $n$. Moreover, quantitative results can be
obtained if correlations between the particles are somehow weak. This means 
that the interaction between the particles is to be neglected, or,
a more realistic possibility, an appropriate mean field theory is chosen.
The latter results in the picture of free quasiparticles moving in the
effective self-consistent field created by the constituents. 
 
In this limit  the temperature $T$ is a well-defined physical 
quantity and all equilibrium characteristics can be found.
For example, for a gas of noninteracting fermions this results in
the famous Fermi-Dirac distribution (FDD) of the occupation numbers:
%\begin{equation}\label{FDd}
$\overline{n}_\alpha =1/(exp [(\varepsilon _\alpha -\mu)/T]+1)~.$
%\end{equation}
There are many real complex systems, such as compound nuclei, rare-earth
atoms, molecules, quantum dots, etc., which do not satisfy
the conditions for FDD to hold. 
The number of active particles in these systems can be relatively small
($\sim 10$), and the interaction between them (even the residual
interaction in the mean-field basis) is large, i.e., greater than the
energy intervals between unperturbed basis states. However,
 this interaction makes up for the absence of a heat bath, and promotes
the onset of ``randomization'' and quantum chaos\cite{Ce,Zel,FIC96}, 
which gives one a possibility to talk about some 
kind of equilibrium in the system, and pursue the development of
a statistical theory for few-body Fermi systems \cite{FIC96}.

%**************************************************************
{\bf 2. The Ce atom.}
The cerium atom has one of the most complicated spectra in the periodic
table.
Its electronic structure consists of a Xe-like
$1s^2\dots 5p^6$ core and four valence electrons. A large difference in the
energy scales of the core and valence electrons allows us to neglect
excitations from the core and consider the wave function of the core as a
``vacuum'' state $|0\rangle$. Accordingly, the four active
electrons  form the spectrum of Ce below the ionization threshold.

The calculations are performed using the Hartree-Fock-Dirac (HFD) and 
configuration interaction (CI) methods (see \cite{Ce} for details).
A self-consistent HFD calculation determines the
mean-field potential, which is then used to calculate the basis set
of single-particle ortho-normalized relativistic states
$|\alpha \rangle = |nljj_z\rangle $ with energies $\varepsilon _\alpha$.
This procedure defines the zeroth-order Hamiltonian of the system,
$\hat{H}^{(0)}=\sum_{\alpha}\varepsilon_\alpha a^\dagger _\alpha
a_\alpha ~.$
The unperturbed multi-particle basis states $|k\rangle $ constructed
from the single-particle states, $|k\rangle = a^\dagger _{\nu_1}
a^\dagger _{\nu_2}a^\dagger _{\nu_3}a^\dagger _{\nu_4}|0\rangle $,
are eigenstates of $\hat H^{(0)}$:
$\hat{H}^{(0)}|k \rangle = E^{(0)}_k|k\rangle $, where
$E^{(0)}_k = \sum _\alpha \varepsilon_{\alpha}n^{(k)}_{\alpha}$
and  $n^{(k)}_{\alpha}=\langle k|a^\dagger _\alpha a_\alpha |k\rangle $ 
are the occupation numbers equal to 0 or 1.

The total Hamiltonian $\hat H$ of the active electrons is the sum of the
zeroth-order mean-field Hamiltonian $\hat H^{(0)}$ and the 2-body residual
interaction is\\
$\widehat{V}=\frac{1}{2}\sum _{\alpha \beta \gamma \delta }
V_{\alpha \beta \gamma \delta }
a^\dagger _\alpha a^\dagger _\beta a_\gamma a_\delta ~$.
This interaction  contributes to the diagonal and
off-diagonal matrix elements between the multi-particle states $|k\rangle $.
The diagonal part represents an additional energy shift of the state $k$,
and can still be expressed in terms of the occupation numbers.
The off-diagonal matrix elements  are responsible for mixing of the
multi-particle basis states.

Complete diagonalization of the operator $\hat{H}=\hat{H_d}+\hat{V}$
within the space of the $|k \rangle $ basis states produces ``exact''
energies $E_i$ and stationary states $|i\rangle $:
\begin{equation}\label{fulldia}
\hat{H} |i\rangle ={\em E_i} |i\rangle ~,\quad
|i\rangle =\sum_k C_k^{(i)}|k\rangle.
\vspace*{-0.25cm}
\end{equation}

 We included 14 relativistic subshells $nlj$ in the
calculation ($6s$, $7s$, $6p$, $7p$, $5d$, $6d$, $4f$, and $5f$), and
performed exact diagonalization of the $N\!\times \!N$ Hamiltonian
matrix in a Hilbert space with $N\sim 8\times 10^3$, obtained by truncating
the complete set of the shell-model atomic configurations.
To subtract additional symmetries our numerical
results are obtained for the even states of Ce with the total angular
momentum projection set to $J_z=0$. 

The relativistic atomic subshells $nlj$ are $g_s=2j\!+\!1$ degenerate,
therefore, we consider average occupation numbers for the subshells $s$:
$\hat n_s=g_s^{-1}\sum _{\alpha \in s}a^\dagger _\alpha a_\alpha $.
When the number of active particles is small, the occupation numbers for
any eigenstate $n_s^{(i)}=\langle i|\hat n_s|i\rangle$ show strong 
level-to-level fluctuations, and it is more instructive to look at
spectrally averaged values
\begin{equation}\label{nsE}
n_s(E)=\overline{\langle i|\hat n_s|i\rangle }=
\sum_k \overline{\left| C_k^{(i)}\right| ^2}\langle k|\hat n_s|k \rangle ~,
\vspace*{-0.25cm}
\end{equation}
where the overline means averaging over the eigenstates $i$ within
some energy interval around energy $E$.

\begin{figure}
\epsfxsize=8 truecm
\epsfysize=4 truecm
\vspace*{-.6cm}
\centerline{\epsffile{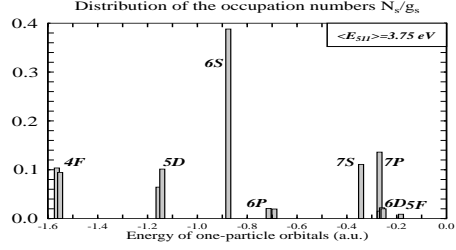}}
\vspace*{-.4cm}
\caption{Occupation numbers $n_s(E)$ (see Eq. (\protect\ref{nsE})),
for the even states of Ce at the excitation energy of $E=3.75$ eV
versus the single-particle energies $\varepsilon _s$ of the orbitals.
\label{nfdsp}}
\vspace*{-.4cm}
\end{figure}

A typical distribution of the occupation numbers calculated at the
excitation energy of 3.75 eV above the atomic ground state is shown in
Fig.~\ref{nfdsp} as a function of the single-particle energy
$\varepsilon _s$ of the orbitals\cite{quas}. One can see that the 
distribution does
not look at all like a monotonically decreasing FDD, as some higher-energy
orbitals have larger $n_s(E)$ than the lower ones.
Moreover, this behaviour persists over the whole energy interval from
the ground state to the ionization potential. For example, the lowest
even state of Ce has a configuration of $4f^26s^2$, while the FDD would
tell us that all 4 electrons must be placed in the lowest $4f$-orbital,
when the energy of the system is low.
At first sight such a strong deviation from the FDD in a strongly
interacting Fermi system speaks against any possibility of statistical
description of the system. However, we show that strongly
interacting orbitals can be properly incorporated in the canonical
ensemble description of the system, and thermally averaged occupation
numbers $n_s(T)$ derived.

%**************************************************************************
{\bf 3. Statistical model.}
Let us  perform a statistical calculation of the occupation numbers
for a system of $n$ particles distributed over $r$ orbitals with energies
$\varepsilon _s$ and degeneracies $g_s$ ($s=1,\dots ,r$). We will assume
that the two-body interaction of any two particles in the orbitals $s$
and $p$ is $U_{sp}$, where both the direct and exchange terms are included:
\begin{equation}\label{Usp}
U_{sp}=\frac{1}{g_s(g_p-\delta _{sp})}
\sum _{\alpha \in s}\sum _{\beta \in p}
(V_{\alpha \beta \beta \alpha }-V_{\alpha \beta \alpha \beta })~.
\vspace*{-0.25cm}
\end{equation}
The energy of a particular many-particle state $k$ is now given by
\begin{equation}\label{Ek}
E_k=\sum _{s=1}^{r}{\cal N}_s\varepsilon _s+\sum _{s=1}^{r}\sum _{p=s}^{r}
\frac{{\cal N}_s({\cal N}_p-\delta _{sp})}{1+\delta _{sp}}U_{sp}~,
\vspace*{-0.25cm}
\end{equation}
where ${\cal N}_s$ is an integer number of particles in the orbital $s$
($0\leq {\cal N}_s\leq g_s$), and $\sum _s{\cal N}_s=n$. The state $k$ is
defined by specifying the orbital occupation numbers ${\cal N}_s$,
and is $G_k$-degenerate, where $G_k=\prod _{s=1}^{r}
\left( {g_s\atop {\cal N}_s}\right)$.

\begin{figure}%[h]
\vspace*{-.5cm}
\epsfxsize=13.0 truecm
\epsfysize=4 truecm
\centerline{\epsffile{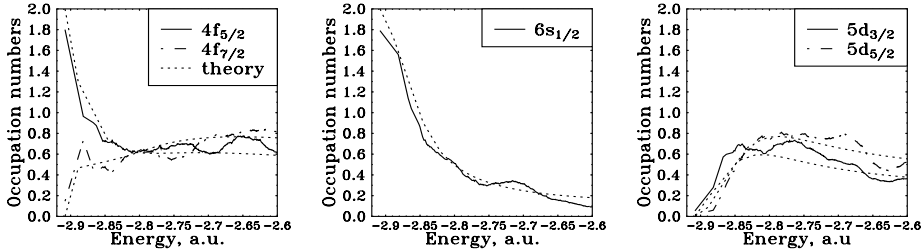}}
\vspace*{-.4cm}
\caption{Comparison of the orbital occupancies $g_s n_s(E)$
obtained from the exact diagonalization (solid and dash-dot 
lines) with $g_sn_s(T({\overline E}))$ obtained from our statistical
theory (dotted lines).\label{compar}}
\vspace*{-0.4cm}
\end{figure}

In the canonical ensemble at temperature $T$ the probability of finding
the system in the state $k$ is given by
\begin{equation}\label{canonw}
w_k=Z^{-1}G_k\exp (-E_k/T)~,
\mbox{ where }\quad Z=\sum _kG_k\exp (-E_k/T)~
\vspace*{-0.25cm}
\end{equation}
and the sum over $k$ runs over all possible multi-particle states
(possibly, with the restriction of parity). The average occupation
numbers $n_s(T)=\overline{{\cal N}_s}/g_s$ are calculated as
\begin{equation}\label{nsT1}
n_s(T)=g_s^{-1}\sum _k{\cal N}_s^{(k)}w_k~,
\vspace*{-0.25cm}
\end{equation}
where ${\cal N}_s^{(k)}$ is the number of particles in the orbital $s$
in the multi-particle state $k$. The energy of the system at a given
temperature is
$\overline{E}(T)=\sum _k E_kw_k $.
 If we know ${\overline E}(T)$, we can find inverse function 
$T({\overline E})$,and  use it in (\ref{nsT1}) in order
to compare with the data from (\ref{nsE}).
 This comparison is shown in figure \ref{compar} and good agreement
is observed.

{\bf 4. Discussion.}
This agreement means that a kind of equilibrium is indeed induced in the
system due to the interaction between the particles (``micro-canonical''
distribution). Moreover, the averaging over it yields results close to
those over a canonical ensemble (\ref{canonw}), with the temperature
chosen to reproduce the total energy of the system. This equivalence is
always true for large systems where any, albeit weak, interaction between
particles leads to equilibrium. However, in a {\em few-particle} system
the residual two-body interaction must be {\em strong} to produce chaotic
eigenstates and facilitate statistical description.

Note, that although the temperature-based description is valid for our
4-particle system, the orbital occupancies could not be described by the 
FDD (Fig. \ref{nfdsp}). The FDD is inapplicable to our system because of
the strong interaction between the particles [second term on the the
right-hand side of Eq. (\ref{Ek})]. However, the deviation from the FDD
is determined not by the magnitude of $U_{sp}$, but rather by the size
of their fluctuations. To see this assume for a moment that
$U_{sp}\equiv U$ are the same. In this case the double sum in
Eq. (\ref{Ek}) just shifts all energies by $\frac{U}{2}N(N-1)$, and
the statistical properties of the system are the same as for
noninteracting particles. If $U_{sp}$ are different for different
orbital pairs $sp$ one can still introduce some average interaction
$\overline{U}$ and subtract this ``background'' interaction from
the interaction term in Eq. (\ref{Ek}). This procedure will effectively
suppress the interaction term, since the summands in expressions like
$\sum_{s<p} (U_{sp}-\overline{U}){\cal N}_p$ have different signs. Note
that the introduction of (energy-dependent) $\overline{U}$ is equivalent
to a mean-field approximation. This approximation is
good if the fluctuations of $U_{sp}$ from one orbital to another are
relatively small. This formulates a condition for the FDD
to be valid. In the Ce atom the situation is just opposite.
The $4f$ orbital has a much smaller radius than any other orbital, hence
the Coulomb interactions $U_{4f4f}$ or $U_{4fs}$ are much greater than
any other $U_{sp}$ (here $s$ and $p$ are orbitals other than $4f$).


\begin{thebibliography}{99}
\bibitem{quas}
 Of course, when this interaction is strong one needs to introduce
some new mean-field orbital energies $\tilde \varepsilon _s$ that would
incorporate the effect of such interaction.
The value $\tilde \varepsilon _s$ will inevitably depend on
the distribution of the other electrons, and hence on the excitation
energy of the system. However, even when $\tilde \varepsilon _s$
are used there is a noticeable deviation from the simple FDD distribution.
\end{thebibliography}
\end{document}